\documentclass{PoS}

\title{Stringy Origin of Discrete R-symmetries}

\ShortTitle{Discrete R-symmetries}

\author{\speaker{Hans Peter Nilles}\\
        Bethe Center for Theoretical Physics and Physikalisches Institut der Universit\"at Bonn, \\ Nussallee 12, 53115 Bonn, Germany\\
        E-mail: \email{nilles@th.physik.uni-bonn.de}}

\abstract{Discrete symmetries play a crucial role in particle physics. 
They appear abundantly in string model constructions. 
We focus here on the case of discrete $R$-symmetries
which are intrinsically connected to the Lorentz group in extra dimensions and the 
appearance of $N$-extended supersymmetry. In that sense, discrete $R$-symmetries
can be understood as "fractionally" extended supersymmetry.
These symmetries reveal insight about the
location of fields in extra dimensions (in particular the Higgs boson). Applications can be
found in the solution of the $\mu$-problem, suppression of proton decay and the structure
of the soft terms of broken supersymmetry.}

\FullConference{Corfu Summer Institute 2016 "School and Workshops on Elementary Particle Physics and Gravity"\\
		31 August - 23 September, 2016\\
		Corfu, Greece}

\begin{document}

\section{Introduction}

Discrete symmetries play a crucial role in fundamental physics. Apart from the basic space-time symmetries C, P and T there are various applications in models of elementary particle physics. Most notably they can provide us with
\begin{itemize}
\item flavour symmetries to explain the pattern of masses and mixing angles of quarks and leptons,

\item hierarchical structure of scales as required by supersymmetry breakdown and the question of the so-called $\mu$-problem,

\item proton-stability and the structure of Baryon($B$)- and Lepton($L$)-number violation,

\item approximate accidental symmetries as e.g. an axionic solution of the strong CP-problem.

\end{itemize}

Discrete symmetries might explain the smallness of some of the parameters in the standard model of particle physics as e.g. the size of the electron mass compared to the masses of other elementary particles. These parameters are forbidden by the symmetries and a "slight" breakdown  can lead to mass hierarchies and small parameters in a natural way. Among the various mechanisms of breakdown, two seem to be of particular significance:

\begin{itemize}

\item  breakdown via non-perturbative effects represented by $\exp{(-X)}$ where a moderately large $X$ can lead to a tiny effect. This could e.g. explain approximate proton stability in the standard model\cite{A} or a small scale of supersymmetry breakdown in the presence of a gaugino condensate\cite{B},

\item the so-called Froggatt-Nielsen\cite{C} mechanism where a small parameter $\epsilon$ is raised to various powers $\epsilon^{n}$ with integer values of $n$. This one seems to be particularly suited for a discussion of quark masses and mixing angles.

\end{itemize}

In bottom-up model building one usually imposes discrete symmetries by hand and studies their phenomenological consequences\cite{D}. This allows us to see the potential importance of some specific symmetries. In a next step one would then combine this with a top-down approach to understand the origin of these symmetries in an ultraviolet (UV) consistent theory. It is this top-down approach which we discuss here. As a UV-completion we shall consider superstring theory in its various versions and analyze the emergence of discrete symmetries. We shall see that the interplay of string-theory selection rules\cite{E} with properties of the compact manifolds (with six space dimensions) can yield a variety of discrete symmetries. We shall describe the emergence of these symmetries in general and shall then focus on a specific class of symmetries: discrete $R$-symmetries. They seem to be intimately connected to the Lorentz group in extra dimensions and the question of the breakdown of
extended supersymmetry. They could provide a solution to the $\mu$-problem in the supersymmetric extension of the standard model, be of relevance for proton stability and provide candidates for dark matter in the universe. They could also be useful for model constructions that might yield examples for so-called "natural supersymmetry"\cite{F}. As a by-product this analysis will shed light on the possible embeddings of the standard model in the framework of string theory. Important ingredients are the geometry of extra dimensions, the "geography" of fields in extra dimensions\cite{FF}
and the appearance of so-called "local" grand unification"\cite{G}.

\section{Strings and particle physics}

Consistent string theories predict ten space-time dimensions and supersymmetry(SUSY). They contain fundamental interactions like gravity and gauge symmetries. When compactified to four space-time dimensions they can provide us with matter multiplets for quarks and leptons. Thus all the ingredients of the (supersymmetric extension) of the standard model can be found within string model building\cite{H}. But string theory gives us more; in particular the discrete symmetries that we want to analyze here.

Unfortunately, the standard model (and its supersymmetric extension (MSSM)\cite{HH})
is not a generic prediction of string theory. In general one has to explore special regions in the "Landscape" of string theory where the gauge symmetries and matter fields of the MSSM appear. It turns out that a rich spectrum of discrete symmetries can be accommodated in these models as well. What is the origin of these discrete symmetries? A first source is the "geometry" of the compact extra dimensions. The compact manifold can have geometrical symmetries that reflect themselves in the low-energy effective 4-dimensional field theory. A second source is the "geography" of fields in extra dimensions. Some of the MSSM fields can be localized at special points (or lower dimensional sub-spaces: 
so-called "branes"\footnote{n-branes sweep an n+1-dimensional world-volume.})
within the compact manifold. This could increase the spectrum of discrete symmetries if fields reside at some specific locations with non-generically enhanced symmetries. The relative location of matter-, Higgs- or gauge-fields determines the properties of low energy physics (e.g. Yukawa couplings).

The specific geographical considerations depend on the type of string theory under consideration. Gravity is usually acting in the full $D=10$-dimensional space-time: i.e. in the bulk of the extra dimensions.
For the other fields we 
have the following basic options\cite{H}:

\begin{itemize}
\item M-theory (type IIA)\\
    gauge fields on 6-branes, matter on 3-branes (at the intersection of 6-branes),
\item F-theory (type IIB)\\
    gauge fields on 7-branes, matter on 5-branes (Yukawa couplings at intersection of 5-branes),

\item Heterotic string\\
    gauge fields in the bulk (with non-trivial profile at various branes), matter fields in bulk, on fixed 
    tori ("5-branes") or  
    on fixed
    points ("3-branes").
\end{itemize}

From the phenomenological point of view we find that, in general, chiral fields prefer a stronger localization: quarks and leptons reside at lower dimensional branes or intersections. Higgs fields come in vector-like pairs and are not protected by chiral symmetries (the origin of the so-called $\mu$-problem). Still one pair needs a special protection and this could come from discrete symmetries. Gauge fields are expected to live on higher-dimensional branes (perhaps with a special profile in extra dimensions to exhibit the mechanism of "local grand unification"\cite{G}). In that sense Higgs- and gauge-fields show a similar behaviour. In some cases the Higgs-fields could emerge from higher-dimensional gauge fields (gauge-Higgs unification).

\section{A specific example (explaining the role of geometry and geography)}

Let us start our discussion with a simple 1-dimensional example, a compactified 5th dimension on a circle\cite{I}. We divide (orbifold) the circle by a $Z_{2}$ reflection symmetry and this leads to an interval with two fixed points at the boundaries as shown in Fig. 1. 

\begin{figure}[h]

\centerline{\includegraphics{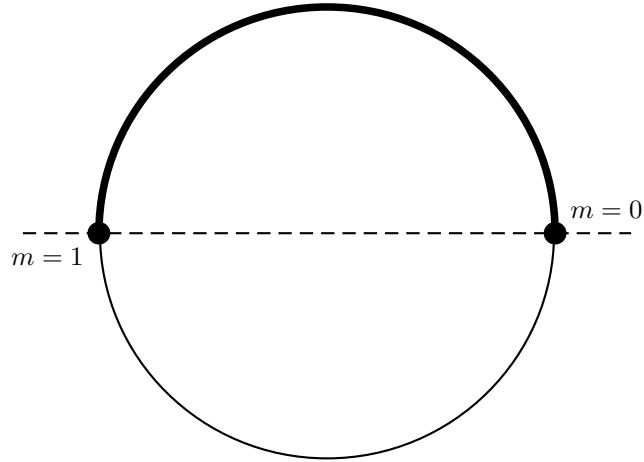}}

\caption{$S_1/Z_2$ orbifold with two fixed points $m=0,1$.} 

\end{figure}
There are two types of fields: 

\begin{itemize}
\item untwisted fields living in the bulk (full 5-dimensional space-time),

\item twisted fields localized at the fixed points (corresponding to "3-branes" in the 5-dimensional theory).
\end{itemize}.

The system under consideration has the following discrete symmetries:

\begin{itemize}

\item an $S_{2}$ symmetry from the interchange of the two fixed points  (a symmetry coming from the geometry of the system),

\item a $Z_{2}\times Z_{2}$ symmetry from string theory selection rules for the twisted fields of the orbifold. With the $Z_{2}$-twist, allowed couplings have to have an even number of the respective twist fields at $m=0$ and 1. This implies a $Z_{2}\times Z_{2}$ symmetry of the correlation functions: 
these are the further symmetries that arise from the "geography" of fields in compactified dimensions.

\end{itemize}
The full discrete symmetry is now a particular combination of the "geometrical"  $S_{2}$ and the 
"geographical" $Z_{2}\times Z_{2}$ symmetry. The twisted fields can be understood as doublets of the $Z_{2}$ symmetries. In this representation the elements of $Z_{2}\times Z_{2}$ are given by 
$\pm 1_{(2x2)}$  and $\pm \sigma_3$ where $1_{(2x2)}$ is the identity
and 
\begin{equation}
\sigma_3=\left(\begin{array}{cc}

  1 &  0 \\

  0 &  -1

 \end{array}\right)
\end{equation}
is the diagonal Pauli matrix.
The symmetry $S_{2}$ interchanges the fixed points and is thus represented by
the Pauli matrix 
\begin{equation}
\sigma_1=\left(\begin{array}{cc}

  0 &  1 \\

  1 &  0

 \end{array}\right).
\end{equation}
The full discrete symmetry is then given by the multiplicative closure of $S_{2}$ and $Z_{2}\times Z_{2}$. It leads to a group with 8 elements:
\begin{equation}
\pm 1,\ \ \ \ \  \pm\sigma_3, \ \ \ \ \ \pm\sigma_1, \ \ \ \ \ \pm i\sigma_2.
\end{equation}
They form the non-abelian group known as $D_{4}$, the symmetry group of the square.
The example shows how discrete symmetries arise from the interplay of geometrical symmetries (here $S_{2}$) and string theory selection rules (here $Z_{2}\times Z_{2}$). Even simple set-ups can lead to large groups (here $S_{2} \rightarrow D_{4}$) and non-abelian discrete groups can arise from "abelian" geometry\cite{I}.

These groups can have immediate applications as flavour symmetries: $D_{4}$ is a non-abelian sub-group of flavour-$SU(2)$ under which the first two generations of quarks
transform as a doublet. The abundance of discrete groups allows to make direct contact to "bottom-up" approaches of flavour physics. For a review see ref. \cite{D}. 

\section{Lessons from the "MiniLandscape"}

This discrete $D_{4}$ flavour-symmetry has been found in explicit string theory constructions as e.g. the orbifold compactification of heterotic string theory\cite{J}. For a review see ref.\cite{K}. This was an attempt to construct string models that could serve as a consistent ultra-violet completion of the MSSM. The successful models allow a study of the "geographical" location of the various fields. This then allows us conclusions about the influence of geometrical properties on the structure of the low-energy effective action. A first lesson of this study was the understanding of discrete flavour symmetries (as $D_{4}$) within a consistent string compactification. But there are further lessons from the MiniLandscape construction\cite{L}. We believe that  these geometrical lessons will be valid beyond the specific theory discussed here and will also apply for other constructions (such as type IIA, IIB or F-theory). As we have already remarked before, chiral representations (relevant for quarks and leptons) tend to have a "stronger" localization. Symmetries like $D_{4}$ appear as a result of fields localized at fixed points in extra dimensions. Gauge fields are different. In the heterotic theory they are bulk fields descending from $E_{8}\times E_{8}$ and this gauge symmetry has to be broken in the process of compactification. Still this leads  to a specific profile of gauge fields in extra dimensions: a phenomenon called "Local Grand Unification"\cite{G}. While in 4 dimensions we have the gauge symmetry $SU(3) \times SU(2)\times U(1)$, various fields (depending on their localization) are subject to higher symmetries like $SU(5)$ or  $SO(10)$ (or other successful grand unified groups). This can explain why some multiplets come in full and some in split representations  of the grand unified group (see ref.\cite{M} for a review). Local grand unification thus preserves the successful aspects of GUTs and avoids some of the more problematic ones, it explains e.g. the doublet-triplet splitting of the Higgs-multiplets.

More lessons of the string constructions can be learned in the study of the Higgs-system. They come in vector-like pairs ($H, \bar{H)}$) and are thus not protected by a chiral symmetry. In the MiniLandscape one aimed at models with 3 families of quarks and leptons. Typical models contained a plethora of vector-like pairs in various  representations of $SU(3)\times SU(2)\times U(1)$. At an intermediate step these pairs were protected by additional $U(1)$ gauge symmetries in 
$E_{8}\times E_{8}$. These symmetries can be broken by vacuum expectation values of singlet fields charged under the additional $U(1)$s. This then allows a mass term for the vector-like pairs and removes them from the low-energy spectrum. We have now to find a mechanism that removes all the vector-like exotics (including additional pairs of Higgs-doublets) and keep exactly one of the Higgs-doublet-pairs light. In the framework of supersymmetric models this is known as the $\mu$-problem. Within the MiniLandscape we can identify models that show a solution of the $\mu$-problem. In explicit constructions we remove vector-like exotics through couplings to singlet fields. This requires a determination of allowed terms in the superpotential 
\begin{equation}
{\phi_1}^{n_1}....... {\phi_k}^{n_k}   H \bar{H}
\end{equation}
at various orders in the power of singlet fields $\phi_i$. Typically most vector-like pairs gain a mass at the first few orders (some-times all the Higgs-pairs are gone). But there are models where a Higgs-pair remains light even at higher orders in the singlet fields\footnote{Due to technical
complications there is a limitation to the power of singlet fields that we can analyze\cite{L}.}. 
It seems that the $\mu$-parameter of this particular Higgs-pair is protected by some new mechanism. A detailed inspection of these models reveals two important facts\cite{L}:
\begin{itemize}
\item the Higgs-pair is living in the bulk, 
\item its $\mu$-parameter is protected by a discrete $R$-symmetry.
\end{itemize}
This is a remarkable observation. The explicit string construction has provided a solution to the $\mu$-problem following from geometric and geographical properties of the compact manifold\cite{N}: the Higgs-pair lives in the bulk! But what is the origin of these discrete $R$-symmetries?

\section{$R$-symmetries}

$R$-symmetries are defined as those symmetries that transform the superspace coordinate $\theta$ non-trivially. They thus do not commute with SUSY, the various components of super-multiplets have different $R$-charges and the superpotential W transforms nontrivially as well. In some ways, $R$-symmetry can be understood as an extension of SUSY as it forbids certain terms that would otherwise be allowed. This has been used as an extension of the MSSM ($N=1$ SUSY) with a $U(1)_{R}$  symmetry to forbid the $\mu$-term. Such a model is not realistic as it would also forbid gaugino masses and trilinear soft A-terms once SUSY is spontaneously broken. To avoid these problems one has imposed  a discrete subgroup of $U(1)_{R}$ (like $Z_{2}$ matter parity). This symmetry avoids proton decay via dimension-4 operators and guarantees that the lightest supersymmetric particle is stable  to serve as a viable candidate for cold dark matter. The discrete $Z_{2}$ matter parity is sometimes called $R$-parity as it is a subgroup of $U(1)_{R}$. It is itself, however, not an $R$-symmetry\footnote{There are no $R$-symmetries of order 2 as the sign of fermion fields can be removed.}. Moreover, $Z_{2}$ - matter parity allows a $\mu$-term and gaugino masses.

When we now turn to the string model constructions we should try to understand the potential origin of $R$-symmetries. They transform bosons and fermions  in a different manner and they share  this property with the Lorentz group. In ten dimensions the Lorentz group is $SO(9,1)$ which can be decomposed in $SO(3,1)\times SO(6)$, so $R$-symmetries should be connected to the remnants of the $SO(6)$ part of the Lorentz group in extra dimensions. Since this group acts on bulk fields in extra dimensions we now understand that the protected Higgs-pairs should be preferably bulk fields
(or should at least have some extension into the bulk). 
This is a rather model independent statement and we think it is true not only in the framework of the heterotic string but of some more general validity.

As stated above, the $R$-symmetries are crucial for the discussion of the $\mu$-problem. This has been discussed in detail in ref.\cite{N} and we shall here only present the general picture. We focus on models of the heterotic MiniLandscape that exhibit the standard model gauge group and the chiral matter content of the MSSM. The first examples were based an the $Z_{6}$-II orbifold \cite{O} and have later been extended to more general cases\cite{P}. The discrete symmetry of the geometry leads to a large number of discrete symmetries governing the couplings of the effective field theory\cite{E}. Apart from the various bosonic discrete symmetries one has for the $Z_{6}$-II orbifold 
a symmetry that could be as big as
\begin{equation}
[Z_6\times Z_3\times Z_2]_R,
\end{equation}
other orbifolds have corresponding discrete $R$-symmetries. As in the case of the bosonic discrete symmetries, the $R$-symmetries of the low energy effective field theory have their origin in a combination of geometry (here the Lorentz group in extra dimensions) and the geography of fields and their string theory selection rules. This opens the possibility for large discrete groups (from geometry and outer automorphisms). The selection rules have been under debate for some time. For the case of orbifolds on factorizable tori (including Wilson lines) they have been clarified in ref.\cite{Q} \footnote{They are not yet understood for models with a freely acting twist.}.

Let us come back now to the specific models of the MiniLandscape with a bulk field Higgs-pair ($H, \bar{H}$). In this case the Higgs-pair $H, \bar{H}$ is neutral under all selection rules\cite{N} while the superpotential transforms nontrivially under the $R$-symmetry. We now consider the superpotential
\begin{equation}
W= P_{\mu}(\phi_{i}) H\bar{H}
\end{equation}
where $P_{\mu}(\phi_{i})$ is a polynomial of (singlet) fields $\phi_{i}$ with some of them transforming nontrivially under the $R$-symmetry. Unbroken $R$-symmetry requires a vanishing vacuum expectation value of $P_{\mu}(\phi_{i})$ and thus a vanishing $\mu$-term. But there are some more interesting consequences
of the $R$-symmetry. Consider
\begin{equation}
W=   P_{W}(\phi_{i})+   P_{\mu}(\phi_{i}) H\bar{H}
\end{equation}
with a polynomial $P_{W}(\phi_{i})$ (transforming nontrivially under the $R$-symmetry). The symmetry forbids a vacuum expectation value of $P_{W}(\phi_{i})$ and this implies that the supersymmetric ground state is flat Minkowski space. Unbroken symmetry corresponds to vanishing gravitino mass and $\mu = 0$. This solution of the $\mu$-problem in the string models is a consequence of remnants of the Lorentz group in higher dimensions. In a bottom-up construction it had been  discussed by Casas and Munoz\cite{R}.

\section{Connections to extended SUSY}

The amount of unbroken SUSY depends on the manifold of compactification (and its holonomy group). The compactification of a $D=10$ dimensional $N=1$ SUSY theory can maximally lead to $N=4$ extended SUSY in $D=4$. This is achieved with a manifold of trivial holonomy. $N=2$ extended SUSY is obtained  for a manifold  of $SU(2)$ holonomy and $N=1$ with manifolds of $SU(3)$ holonomy (Calabi-Yau manifolds). Within brane-world models and orbifold compactifications we are typically dealing with discrete subgroups of the holonomy groups mentioned above. This results in (extended) SUSY with additional discrete symmetries. The appearance of extended SUSY is thus strongly correlated with the geometry of compactified space.

Let us look at $N$-extended SUSY in more detail. $N$ denotes the number of superspace coordinates $\theta_{\alpha} (\alpha = 1,2)$ and this determines the structure of the superfields. For $N=1$ we have chiral superfields
\begin{equation}
\phi(x,\theta)=\varphi(x)+\theta\psi(x) +\theta\theta F(x)
\end{equation}
with components $\varphi(x), \psi(x), F(x)$ where $F(x)$ is an auxiliary field. 
Chiral fermions can be described by chiral superfields. 
Gauge symmetries are described by a vector superfield with components $A_{\mu}(x), \lambda(x), D(x)$ (in the Wess-Zumino gauge) where $\lambda$ denotes the gaugino and D is auxiliary.
 Massive gauge fields in $N=1$ are described by bigger multiplets that combine a vector superfield with a chiral superfield in the adjoint representation: $\psi (x)$ and $\lambda (x)$ are combined to a Dirac fermion and $A_{\mu}, \varphi$ combine to a massive gauge boson and a real scalar field (a realization of the supersymmetric Higgs-mechanism). 

For $N=2$ extended SUSY we obtain larger multiplets. The gauge multiplet (at mass zero) has the same degrees of freedom as the massive $N=1$ gauge multiplet. In $N=1$ language it decomposes into a vector and chiral multiplet. In addition we can have so-called hyper-multiplets that decompose as chiral + antichiral multiplets in $N=1$ SUSY. Thus $N=2$ extended SUSY does not have chiral representations. The theory contains an additional  $SU(2)_{R}$  symmetry acting on the superspace coordinates $(\theta_{\alpha})_{1}$ and $(\theta_{\alpha})_{2}$: they transform as a doublet under $SU(2)_{R}$ and since the superspace coordinates transform nontrivially, this an $R$-symmetry. For the $N=2$-vector supermultiplet the scalar and the gauge boson transform as a singlet under $SU(2)_{R}$  while the Weyl fermions $\psi$ and $\lambda$ form a doublet. In the hyper-multiplet the fermions transform as singlets and the pair of complex scalars $\varphi, \bar{\varphi}$ are members of a doublet. So $N$-extended supersymmetry always comes with an $R$-symmetry.

For $N=4$ we have a single superfield in the adjoint representation of the gauge group. It consists of a gauge boson, 4 Weyl fermions and 6 (real) scalars. In $N=2$ language it decomposes into a 
vector-multiplet and a hyper-multiplet. Here the $R$-symmetry is $SU(4)_{R}$: the gauge boson is a singlet and the Weyl fermions form a 4-dimensional representation. The scalars transform as the antisymmetric combination  of $4\times 4$, i.e. a 6-dimensional representation  of $SU(4)_{R}$.

This now allows a direct connection to the symmetry of extra-dimensional space. The algebra of 
the $SO(6)$  subgroup of the Lorentz-group $SO(9,1)$ is isomorphic to the algebra of $SU(4)$. The scalars transforms under the (vector) 6-dimensional representation and the spinors under the 4-dimensional spinor representation of $SO(6)$. The holonomy group is a subgroup of $SO(6) \sim SU(4)_{R}$. We can now discuss the different sectors of our compactified models. Compactification on a 6-torus leads to $N=4$ in $D=4$ with maximal $R$-symmetry $SU(4)_{R}$ of completely geometric origin. The bulk fields transform nontrivial under this full $R$-symmetry. If we consider 5-branes (fixed tori with ($D=6$)) we would find $N=2$ extended supersymmetry with $R$-symmetry up to $SU(2)_{R}$ (partially from geometry, partially from selection rules). Fields on these 5-branes are subject to nontrivial $SU(2)_{R}$ transformations. Fields on 3-branes (fixed points with $D=4$) feel $N=1$ supersymmetry with no $R$-symmetry of geometrical origin. Various sectors of a given model can thus feel different amounts of supersymmetry and $R$-symmetry. This can have important consequences for phenomenological applications\cite{S}. Bulk fields are better protected by these extended symmetries and this is at the origin of the solution of the $\mu$-problem for the Higgs-multiplets living in the bulk. The above discussion shows that there is an intimate relation of SUSY-breakdown and $R$-symmetry breakdown, discussed already in ref.\cite{T}. Given a certain amount of SUSY, additional (discrete) $R$-symmetries improve the properties of the theory. In that sense discrete $R$-symmetries can be understood as a "fractional" extension of supersymmetry.

\section{Applications}

Both, discrete $R$- and non-$R$-symmetries can play crucial roles in model constructions. They are intimately related to the geometric  structure of compactified space. In addition the "geography" of fields in extra dimensions gives rise to many possibilities for discrete groups (through the influence of the string theory selection rules). In many cases the discrete symmetries lead to "better behaved" quantum field theories in the ultraviolet. They might be used to forbid certain counter terms and explain the appearance of large mass hierarchies.

\subsection{Flavour symmetries}

An obvious application concerns the flavour symmetries of quarks and leptons and the pattern of Yukawa couplings. In our simple example we had a non-abelian $D_{4}$ flavour symmetry (as a subgroup of flavour $SU(2)$) where the first two generations of quarks  transform as a doublet. More general symmetries within the orbifold compactification of the heterotic string 
and other string constructions  can be found 
in refs.\cite{I,ACK,ACK1,ACK2}\footnote{For a discussion on non-abelian discrete R-symmetries see ref.\cite{TR}.}. 
These symmetries can be used as building blocks for a more complete description of properties of quarks and leptons. Then one could try to make the connection to bottom-up 
constructions\cite{D}.

\subsection{Suppressed couplings}

The standard model exhibits accidental symmetries at the level of renormalizable couplings connected to baryon ($B$) and lepton ($L$) number. Higher dimensional operators (of dimension 5 and 6) as well as non-perturbative effects tend to violate these accidental symmetries. This is particularly relevant for the supersymmetric extensions of the standard model where e.g. operators of dimension 4 and 5 can lead to proton decay at unacceptable rates. Again, discrete symmetries can come to rescue, most prominently the $Z_{2}$ matter parity of the MSSM. It forbids proton decay via dimension 4 operators. In addition it predicts the presence of a stable particle (the lightest supersymmetric particle (LSP)), a candidate for cold dark matter of the universe. Generalizations have been discussed in form of baryon triality $B_{3}$\cite{U} or proton hexality $H_{6}$\cite{V} to further elucidate the structure of the baryon and lepton sectors of the model. This also concerns the question of lepton-number violating neutrino masses via dimension 5 operators. Discrete $R$-symmetries could play a crucial role in this discussion as well. An example can be found in the recently discussed $Z^{R}_{4}$ symmetry\cite{W} (that contains $Z_{2}$ matter parity as a subgroup). Again these groups could be used as the building blocks of a more complete description of supersymmetric extensions of the MSSM.

\subsection{The $\mu$-problem and SUSY breakdown}

We have already discussed the solution of the $\mu$-problem through $R$-symmetries and the appearance of flat Minkowski space for unbroken supersymmetry: the $R$-symmetry avoids deep supersymmetric vacua in Anti-de Sitter space (AdS). The breakdown of $R$-symmetry and supersymmetry are closely connected as the superpotential transforms non-trivially under both. $R$-symmetries could be broken by non-perturbative effects and this gives an explicit relation to supersymmetry breakdown via hidden sector gaugino condensation\cite{X1,X2}.

\subsection{Natural Supersymmetry}

Extended $R$-symmetries imply a better UV-protection than just supersymmetry alone. The string models discussed here exhibit different sectors that feel different amounts of $R$-symmetry and supersymmetry. The difference is closely connected to the location of fields in extra-dimensional space. Bulk fields (untwisted sector) feel the full $N=4$ supersymmetry at tree level. The sectors of twisted tori might be subject to $N=2$  supersymmetry while localized fields (on fixed points in extra dimensions) live in $N=1$  supersymmetric sectors. Of course, the full model will only have $N=1$ supersymmetry and the sectors with $N=2,4$  will  feel a reduced amount of SUSY because of loop effects through couplings to the $N=1$ sector. This structure of the sectors is similar to the picture of local grand unification where some sectors feel enhanced gauge groups like $SU(5)$ or $SO(10)$ although the full model has only $SU(3)\times SU(2)\times U(1)$ as unbroken gauge group.

Once we discuss the spontaneous breakdown of supersymmetry, the hierarchical structure of the various sectors remains intact and this reflects itself in the pattern of soft SUSY breaking terms. More supersymmetry in a given sector results in a better protection. Thus we obtain hierarchical structures of the soft terms in a natural way. This leads to a pattern of soft terms reminiscent of a mediation 
scheme\cite{Y1} known as mirage mediation\cite{Y2}. 
Bulk fields with $N=4$ will be subject to a no-scale structure 
\cite{YYY}
of the soft terms: they vanish at tree level and will only become nonzero when coupling to other sectors with less supersymmety. Non-trivial $R$-symmetries can protect soft terms as e.g. gaugino masses. Fields localized at fixed points feel the full breakdown of superymmety, they are thus the heaviest of the supersymmetric partners. In the models at hand, these are the (localized) scalars of the first two generations of quarks and leptons. A hierarchical structure of soft terms emerges naturally\cite{S,Z}. This implies that even in the framework of so-called low-scale SUSY only few fields might be found 
 in the TeV-region, while most of them are heavier and outside the reach of accelerators  like the LHC at CERN. The pattern of soft SUSY breaking terms in various sectors of extended SUSY leads to a reduced fine tuning of the parameters.

\section{Conclusions}

Discrete symmetries are crucial ingredients for models of elementary particle physics. Traditional bosonic discrete symmetries could  be relevant in particular for
\begin{itemize}
\item flavour symmetries,
\item hierarchies of Yukawa couplings,
\item an accidental Peccei-Quinn symmetry\cite{AA}.
\end{itemize}
Discrete $R$-symmetries might be important for
\begin{itemize}
\item a solution of $\mu$-problem in theories with Minkowski vacuum,
\item the question of proton stability and the stability of the LSP,
\item pattern of soft terms for "Natural SUSY".
\end{itemize}
The origin of discrete symmetries can be found in the geometry of compactified space and the location (geography)  of fields in extra dimensions. They appear as a combination of symmetries of the compact manifold and string theory selection rules. Large symmetries can arise from small building blocks.

\section*{Acknowledgements}
I would like to thank Rolf Kappl for enlightening discussions. This work was supported by the
SFB-Transregio TR33 "The Dark Universe" by Deutsche Forschungsgemeinschaft (DFG).

\end{document}